# Reduced Order Modeling of Energetic Materials Using Physics-Aware Recurrent Convolutional Neural Networks in a Latent Space (LatentPARC)

Zoë J. Gray[a], Joseph B. Choi[a], Youngsoo Choi[b], H. Keo Springer[b], H. S. Udaykumar[c], Stephen S. Baek*,[a],[d]

**Abstract:** Physics-aware deep learning (PADL) has gained popularity for use in complex spatiotemporal dynamics (field evolution) simulations, such as those that arise frequently in computational modeling of energetic materials (EM). Here, we show that the challenge PADL methods face while learning complex field evolution problems can be simplified and accelerated by decoupling it into two tasks: learning complex geometric features in evolving fields and modeling dynamics over these features in a lower dimensional feature space. To accomplish this, we build upon our previous work on physics-aware recurrent convolutions (PARC). PARC embeds knowledge of underlying physics into its neural network architecture for more robust and accurate prediction of evolving physical fields. PARC was shown to effectively learn complex nonlinear features such as the formation of hotspots and coupled shock fronts in various initiation scenarios of EMs, as a function of microstructures, serving effectively as a microstructure-aware burn model. In this work, we further accelerate PARC and reduce its computational cost by projecting the original dynamics onto a lower-dimensional invariant manifold, or 'latent space.' The projected latent representation encodes the complex geometry of evolving fields (e.g., temperature and pressure) in a set of data-driven features. The reduced dimension of this latent space allows us to learn the dynamics during the initiation of EM with a lighter and more efficient model. We observe a significant decrease in training and inference time while maintaining results comparable to PARC at inference. This work takes steps towards enabling rapid prediction of EM thermomechanics at larger scales and characterization of EM structure-property-performance linkages at a full application scale.

## 1 Introduction

Understanding the relationship between microstructural morphology and sensitivity [1], [2] of energetic materials (EM) is of critical importance as it is directly concerned with the safety of EMs. However, the search space for this problem (i.e., the space of microstructural morphology) is vast. However, experimental methods to study the structure-property-performance (SPP) linkages of EMs are costly and time-consuming, making numerical simulation an attractive alternative. Numerical simulation methods [3], [4], [5], [6], [7], [8] use known physics equations (burn models) and rigorous numerical methods to solve those equations. However, numerical simulation is also prohibitively expensive as the complexity of EM thermomechanics demands a finely resolved simulation grid in space and time, resulting in a single simulation taking on the scale of hours to generate.

To achieve faster simulation, deep learning has recently been used for surrogate models which can train on numerical simulation data to then generate predicted simulations rapidly and accurately at inference. Physics-aware deep learning (PADL) methods build on the recent success of deep learning by imbuing physics knowledge into deep learning models to meet the rigorous scientific requirements in physics research. Additionally, PADL methods can learn from much smaller datasets, a necessity in EM research which is sparse in data when compared to typical deep learning applications. Beyond the EM community, various PADL models have been shown to successfully model complex spatiotemporal dynamics [9], [10], [11], [12], [13], [14]. Of these, physics-aware recurrent neural networks (PARC) have shown particular promise for challenging nonlinear problems characterized by fast transients, sharp gradients, and discontinuities [10], [11], [15], [16]. Especially regarding EMs, once trained, PARC is capable of rapidly and accurately predicting the formation and growth of hotspots during various initiation scenarios of EMs and characterizing EM

*[a] Z.J. Gray, J.B. Choi, S. Baek*
*School of Data Science*
*University of Virginia*
*Charlottesville, VA 22903, United States*

*[b] Y. Choi, H.K. Springer*
*Lawrence Livermore National Laboratory*
*Livermore, CA 94550, United States*

*[c] H.S. Udaykumar*
*Department of Mechanical Engineering*
*University of Iowa*
*Iowa City, IA 52242, United States*

*[d] S. Baek*
*Department of Mechanical and Aerospace Engineering*
*University of Virginia*
*Charlottesville, VA 22903, United States*

** E-mail: baek@virginia.edu*

sensitivity (e.g., James sensitivity envelope) as a function of microstructural morphology [17].

However, even state-of-the-art approaches such as PARC, though achieving speed-ups orders of magnitude faster than numerical simulation methods [10], [11], remain computationally heavy and slower than desired for large-scale exploration of the space of EM morphologies. Recent work has shown that projecting high-dimensional data into learned latent spaces and then subsequently using a PADL method to learn dynamics in the latent space can accelerate surrogate modeling while preserving prediction accuracy [18], [19]. This strategy has demonstrated success in modeling relatively simpler physical systems, but its effectiveness is less established for problems that exhibit sharp gradients, strong nonlinearities, and fast transients characteristic of EM thermomechanics. PARC has not yet been tested in a latent space of compressed dynamical data in prior literature. However, if the ability of PARC models to predict well on EM data can be preserved when trained in a latent space, it could be a powerful tool to further accelerate EM surrogate models.

In this paper, we introduce LatentPARC, a surrogate model for dynamical systems that consists of a PARC model trained in a learned latent space. We match the predictive performance of PARC models trained on non-compressed data while achieving significant reductions in train time, inference time, and model size. This establishes, for the first time, that PARC models can effectively learn dynamics on a reduced-dimensional data manifold. To contextualize our work, we begin with a brief review of data driven surrogate models for modeling dynamical systems for EM. We also review relevant literature on the prior work of Nguyen et al. [10], [11], which demonstrates the ability of PARC models to accurately model EM thermo-mechanics such as hot spot formation and hotspot growth.

# 2 Background

## 2.1 Reduced Order Modeling of EMs

Reduced-order models (ROM) use data from high-fidelity, full-order models (FOM) to construct surrogate models capable of generating new predictions with a reduced computational cost. Among ROM strategies for physics modeling, data-driven approaches, which learn directly from data without requiring explicit knowledge of governing equations, are especially useful when these equations are unknown or intractable. Research on ROMs for modeling EM thermomechanics is limited. Additionally, much of the existing work is focused on molecular dynamics simulations of EM systems, which, while valuable, address a fundamentally different scale and present different challenges than the mesoscale and macroscale thermomechanics of EMs considered here.

We now focus on a subset of ROMs that learn on compressed, lower-dimensional representations of numerical simulation data. ROMs of this type have been applied to problems such as fluid dynamics [19], [20], [21], [22] and material fracture [19]. These problems, though complex, often involve governing equations and boundary conditions that are more idealized and, in certain respects, simpler than those arising in EM initiation problems. Latent space PADL models for EM have been explored in [23] and [24], which demonstrate that reduced representations can effectively capture complex pore collapse dynamics while providing substantial computational speed-ups. Our work is complementary, but distinct, in that we embed a PARC model into a latent space using an encoding method compatible with PARC and use non-idealized pore geometries.

The motivation behind LatentPARC is the manifold hypothesis [25], [26], which suggests that high-dimensional data, despite its complexity, often lies on or near a lower-dimensional manifold embedded within that high-dimensional space. Many methods exist to project data into a reduced-dimensional space and more efficiently capture the essential structure of the original data. Common approaches include autoencoders [27], proper orthogonal decomposition (POD), dynamic mode decomposition (DMD), diffusion maps [28], and more. For our problem, we employ a convolutional autoencoder due to its computational efficiency, inherent spatial and geometric awareness, and ability to reconstruct the original high-dimensional fields from the latent representation.

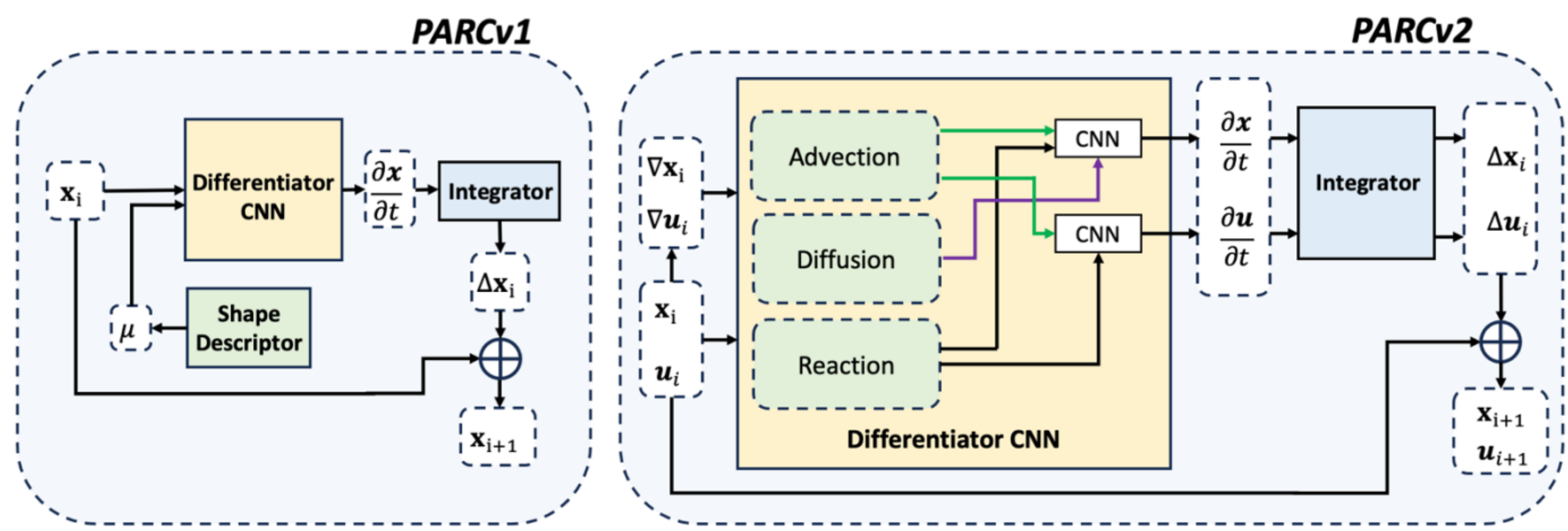

**Figure 1: PARCv1 and PARCv2 architectures.**

## 2.2 Physics Aware Recurrent Convolutional Neural Networks (PARC)

It has previously been demonstrated that PARC models can reliably learn the thermomechanics of EMs across various initiation scenarios as a function of morphology [10], [11], [17], [29]. PARC models use convolutional neural networks (CNNs), which are designed to capture geometric and spatial information, to model the dynamics of systems with nonidealized pore geometrics, something that has not been done with previous methods [24]. The physics awareness of PARC also enables the model to learn effectively from the relatively small number of numerical simulation datasets available for EMs, while also improving convergence and overall predictive performance. It should be mentioned that, unlike some PADL surrogate models, PARC models do not explicitly enforce physical constraints such as conservation laws or laws of shock dynamics. Instead, they rely on architectural inductive bias and patterns learned from the data.

Like traditional numerical solvers for differential equations, the PARC architecture first learns an approximation of the time derivative and then advances the state through integration. Although numerical solvers differ in how they evaluate derivatives, the essential structure is the same: estimate the local time derivative and integrate it forward. PARC models follow this general structure by replacing the derivative evaluation with a neural network that learns to approximate the time derivative. This stands in contrast to many PADL approaches that predict the future state directly from the current one in a single feed-forward mapping, without an explicit notion of integration. Learning to approximate the derivative rather than the full future state is an easier optimization problem and leads to more stable, accurate, and data-efficient models and is discussed more in depth in [11].

$$\frac{\partial \mathbf{x}}{\partial t} = f(\mathbf{x}, \mu \,|\theta) + \varepsilon \tag{1}$$

$$\mathbf{x}(t + \Delta t) = \mathbf{x}(t) + \Delta \boldsymbol{x}\,, \qquad \Delta \boldsymbol{x} = \int_0^{\Delta t} f(\mathbf{x}, \mu \,|\theta) dt \tag{2}$$

This general structure is described by Equations 1 and 2. In Equation 1, the rate of change of the state $\mathbf{x} \in X$ with respect to time, where $X$ is the phase space of the evolving physical fields of temperature and pressure, is a function of the current state $\mathbf{x}$ and the morphology $\mu$. Building from [30], which provides the first universal approximation theorem for operator approximation using neural networks, [31] extends this theorem to more general Banach spaces. According to this universal approximation theorem, the function $f$ can be accurately approximated by a sufficiently deep neural network parameterized by the model parameters $\theta$ and up to some stochastic noise and prediction error $\varepsilon$. Then the approximated $\frac{\partial \mathbf{x}}{\partial t}$ is integrated, as shown in Equation 2, to compute the state update $\Delta\mathbf{x}$ over a time step $\Delta t$. The state update is then added to the current state $\boldsymbol{x}(t)$ to obtain the next state $\mathbf{x}(t + \Delta t)$. Once $f$ is learned, the entire dynamical sequence can be predicted in this manner from just the initial state $\boldsymbol{x}_0$. This general structure is reflected in the architectures of PARCv1 and PARCv2, the two most established PARC models, shown in Figure 1. Here, the differentiator CNN, which learns $f$, is shown in yellow and the integrator is shown in blue.

PARCv1 was shown to predict the response of shocked microstructures within a 95% confidence band of numerical simulation results, while reducing computation time from hours or days (for numerical simulation methods) to less than a second [10]. PARCv1 has also been applied to shock-to-detonation transition (SDT) simulations, where it was used to obtain a sensitivity envelope, with high accuracy when verified against an experimentally obtained sensitivity envelop, in just a few hours of work [17].

Building on this foundation, PARCv2 [11] retains the overall architectural structure of its predecessor while improving the approximation made by the differentiator by providing additional physics-based inputs to the neural network during training: numerically calculated diffusion and advection fields. This addition helps constrain the search space towards more physically consistent outputs. PARCv2 has been evaluated across a range of canonical benchmark problems in fluid dynamics, supersonic flow around cylindrical obstacles, shock-induced pore-collapse in EM, and shear band formation in EM under weak shock loading [11], [23]. Across these applications, PARCv2 has consistently outperformed other PADL methods, including PARCv1. The tradeoff, however, is that PARCv2 is substantially heavier, requiring more time to train and having slower inference time than PARCv1.

# 3 Methods

## 3.1 LatentPARC Architecture

The LatentPARC architecture, shown in Figure 2, consists of an encoding module, pictured in green, coupled with a dynamics module adapted from the original PARCv1 architecture, pictured in yellow and blue. We assume that there exists a mapping $E : X \to Z$ such that the temporal continuity in the original space is preserved in the latent space. Here, $\mathbf{x} \in X$ represents the original high-dimensional physical field data, $\mathbf{z} \in Z$ the lower-dimensional latent representation, and $E$ the function which maps elements in $X$ to $Z$ which corresponds to the encoder portion of the autoencoder [32]. It is also essential that there exists a mapping $D : Z \to X$ so that, after predictions are made in the reduced dimensional $Z$ space, they can be projected back to the original space $X$, where measurements and observations can be made. Here, $D$ corresponds to the decoder component of the autoencoder.

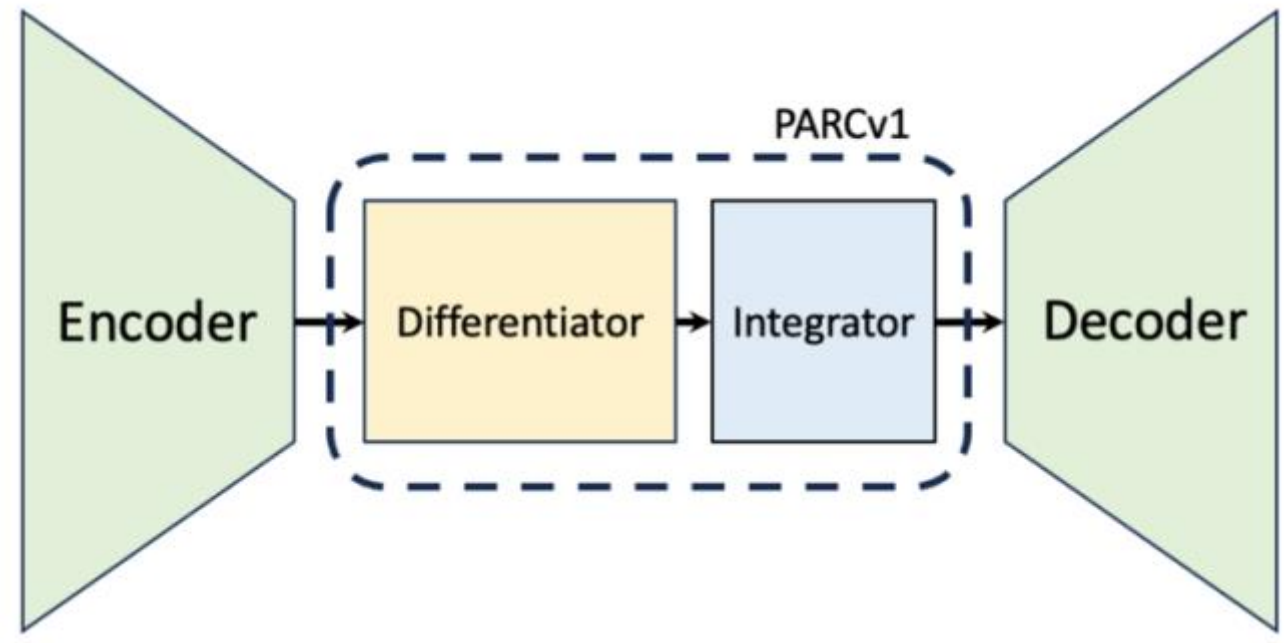

**Figure 2: LatentPARC architecture.**

We use a convolutional autoencoder [32] as the latent encoding module due to their ability to capture key geometric feature information, preserve spatial structure within the latent space, and offer improved parameter efficiency compared to fully connected networks. The ability to capture geometric information and preserve an image-like structure in the latent space is ideal for maximizing the performance of the PARC module, given that PARC is a computer vision model and can take advantage of the rich spatial information contained in image data. Furthermore, the efficiency of convolutional neural networks aligns with our goals of reducing computational cost and memory usage.

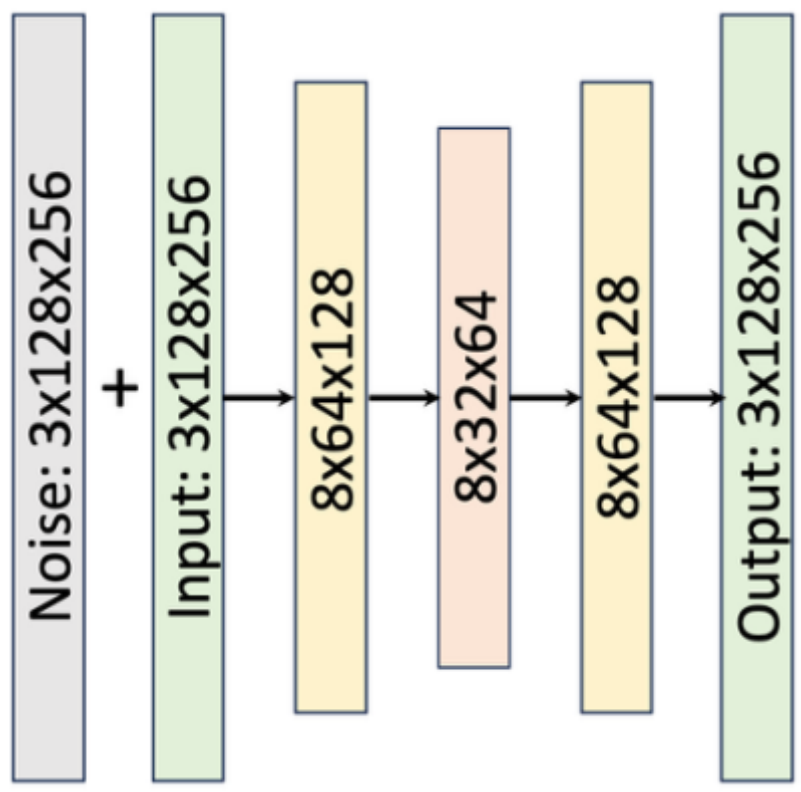

**Figure 3: LatentPARC convolutional autoencoder.**

The exact dimensions of the autoencoder used in this application of LatentPARC is shown in Figure 3. The encoder consists of a series of strided convolutional layers [33] and ReLU activations [34], progressively reducing the spatial resolution of image height and width by 2x, while simultaneously increasing channel depth, with a bottleneck layer acting as the low-dimensional latent space. The decoder mirrors the structure of the encoder, using bilinear up-sampling followed by convolutional layers to work back up to the original data dimension from the latent dimension. Bilinear up-sampling was chosen to avoid artifacts associated with the alternative option of transpose convolutions.

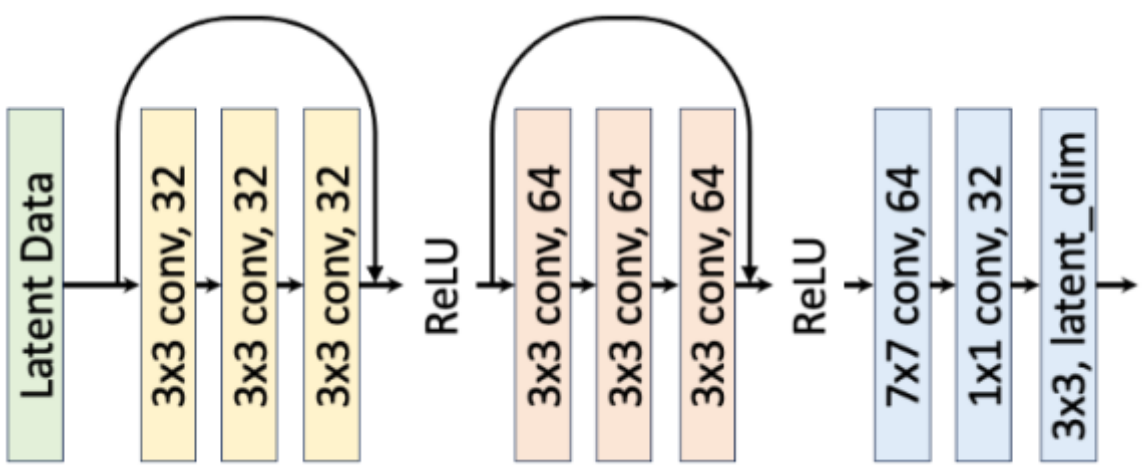

**Figure 4: LatentPARC differentiator.**

The differentiator CNN, detailed in Figure 4, is composed of two ResNet blocks [35], each containing three $3 \times 3$ convolutional layers with skip connections and ReLU activations, designed to capture hierarchical spatial dependencies in the latent representation. This is followed by a $7 \times 7$ convolution layer to expand the receptive field and aggregate context, and a $1 \times 1$ convolution for feature channel mixing and dimensionality compression. A final $3 \times 3$ convolution layer, inspired by super-resolution, is used to enhance the details [36] of the output of the differentiator, $\frac{\partial \mathbf{z}}{\partial t}$. ReLU activations and dropout are used throughout for regularization and to encourage stable learning of latent dynamics.

The dynamics module is a very simple version of a PARC model which more closely resembles the PARCv1 architecture [10], rather than PARCv2 [11], despite PARCv2 being the more capable model. This is because classical advection and diffusion calculations, which are used in the PARCv2 differentiator, are not trivial to calculate in latent coordinates. For example, the calculation of advection requires information about the velocity fields in the x and y directions. However, encoding these fields would lose their physically interpretable structure after transformation into the latent space.

Unlike PARCv1, however, our latent-space formulation does not require a separate shape descriptor. The shape descriptor in PARCv1 is designed to extract features from the morphology; however, since the geometric features of the morphology are embedded directly into the latent fields alongside temperature and pressure via the convolutional autoencoder in LatentPARC, we can do away with this feature.

Furthermore, based on prior empirical results, we use a numerical Runge–Kutta integrator (RK4) [37], which has been shown to significantly improve PARC model's prediction accuracy for hotspot formation and nonlinear temporal dynamics when compared to a CNN-based integrator, which had been experimented with in previous work. First, we let an initial value problem be specified as in Equation 3. Here, $f$ is the differentiator, our unknown function we would like to approximate.

$$\frac{\partial \mathbf{z}}{\partial t} = f(t, \mathbf{z}), \qquad \mathbf{z}(t_0) = \mathbf{z}_0 \tag{3}$$

We then pick a constant step size of 1, since the $\Delta t$ between each data frame is constant across all time steps and simulations. This then results in the RK4 approximation of the future latent state $\mathbf{z}_{n+1}$ being defined by Equation 4, where $\boldsymbol{z}_n$ is the current latent state

$$\boldsymbol{z}_{n+1} = \boldsymbol{z}_n + \frac{1}{6}(k_1 + 2k_2 + 2k_3 + k_4) \tag{4}$$

with $t_{n+1} = t_n + 1$ and

$$k_1 = f(t_n, \boldsymbol{z}_n), \tag{4.1}$$

$$k_2 = f\left(t_n + \frac{1}{2}, \boldsymbol{z}_n + \frac{k_1}{2}\right), \tag{4.2}$$

$$k_3 = f\left(t_n + \frac{1}{2}, \boldsymbol{z}_n + \frac{k_2}{2}\right), \tag{4.3}$$

$$k_4 = f(t_n + 1, \boldsymbol{z}_n + k_3). \tag{4.4}$$

To summarize, the dynamics module of LatentPARC is composed simply of a residual neural network that approximates the latent space time derivative $\frac{\partial \mathbf{z}}{\partial t}$, followed by the RK4 scheme to compute the latent state update $\Delta\mathbf{z}$. The latent state $\boldsymbol{z}_n$ is subsequently updated as shown in Equation 5.

$$\mathbf{z}_{n+1} = \mathbf{z}_n + \Delta\mathbf{z}, \qquad \Delta\boldsymbol{z} = \int_{t_n}^{t_{n+1}} \frac{\partial \mathbf{z}}{\partial t} dt \tag{5}$$

## 3.2 Training the Model

$$L_R = \frac{1}{n}\sum_{t=1}^{n} |\mathbf{x}_t - \bar{\mathbf{x}}_t| \tag{6}$$

$$L_D = \frac{1}{n}\sum_{t=1}^{n} |\mathbf{z}_{t+1} - \hat{\mathbf{z}}_{t+1}| + \frac{1}{n}\sum_{t=1}^{n} |\mathbf{x}_{t+1} - \hat{\mathbf{x}}_{t+1}| \tag{7}$$

The training of the autoencoder and differentiator components is performed separately. We first train the autoencoder to convergence using the reconstruction loss formula, denoted $L_R$ in Equation 6. Here, $\boldsymbol{x}_t$ denotes the synthetic ground truth data and $\bar{\boldsymbol{x}}_t$ denotes the reconstruction, or output of the decoder, at time step $t$. To mitigate over smoothing and encourage the preservation of fine-scale features, we inject additive uniform random noise into the input during training, the process of which is specified in Equation 8. Here, $N$ is the maximum possible magnitude of noise which can be sampled from the random uniform distribution. During our training process, we used a value of $N = 0.16$ which was reduced by a factor of $0.8$x each 500 epochs.

$$\boldsymbol{x}_{noise} = \boldsymbol{x} + \epsilon, \qquad \epsilon \sim Uniform(0, N) \tag{8}$$

This encourages the model to first learn features with larger magnitudes, since lower magnitude features will be obscured by noise, and then learn the finer features as noise is reduced. The addition of this regimen addresses problems specific to the EM dataset which is heterogeneous across channels (microstructure is binary, pressure is relatively normal, and temperature is very skewed). Denoising also helps improve generalization to unseen samples, an additional advantage given our relatively small training dataset [38]. We used an Adam optimizer, a batch size of 32, and a step function learning rate scheduler which starts at a learning rate of $1 \times 10^{-3}$ and decreases by a factor of $0.8$ every $200$ epochs. We trained the autoencoder for $1{,}500$ epochs.

After training the autoencoder independently, its weights are loaded into the LatentPARC architecture and kept fixed throughout the subsequent training phase. The dynamics module—specifically the differentiator component—is then trained using the dynamics loss function $L_D$, as defined in Equation 7. Here, values with a hat denote a prediction of the dynamics module. Specifically, $\hat{\boldsymbol{z}}_{t+1}$ is the prediction of the next time step within the latent space and $\hat{\boldsymbol{x}}_{t+1}$ is the corresponding decoded version of $\hat{\boldsymbol{z}}_{i+1}$. The first term in $L_D$ penalizes errors in the predicted latent state at the next time step, encouraging accurate temporal evolution. The second term acts as a regularization constraint, ensuring that the latent trajectories remain within a subspace that can be reliably decoded by the frozen decoder, thereby preserving consistency with the original data manifold.

To train the latent dynamics module, we implemented a rollout training strategy in which the model predicts a sequence of user specified $n_{ts}$ future latent states in an auto-regressive manner. Starting from an initial latent encoding $\mathbf{z}_t$, the differentiator and numerical integrator are used iteratively to produce $\mathbf{z}_{t+1}, \dots, \mathbf{z}_{t+n_{ts}-1}$. At each time step, the predicted latent state is compared against the corresponding synthetic ground truth using the dynamics loss $L_D$, and the total loss is computed as the average over the full rollout horizon.

This approach allows the model to better capture long-term dependencies, rather than optimizing for one-step prediction accuracy alone. Rollout training is particularly helpful in physical systems where error accumulation over time can lead to unstable or nonphysical behavior. It is also particularly beneficial for physical systems that evolve through distinct phases, each characterized by qualitatively different dynamics, with rapid transitions between them. When training this module, we used an Adam optimizer, batch size of 1, constant learning rate of $1 \times 10^{-5}$, a rollout window of $n_{ts} = 5$, and trained for $3{,}000$ epochs.

## 3.3 Latent Rollout Prediction

Once LatentPARC is fully trained, predictions can then be made entirely within the latent space. This is what allows for the speedup in inference. To make a prediction with LatentPARC, the initial condition $\boldsymbol{x}_0$, composed of the initial temperature and pressure fields and initial pore geometry, is first encoded by the encoder. Then $\boldsymbol{z}_0$, the latent representation of the initial condition, is input to the dynamics module to predict the full sequence of the reaction in an autoregressive manner. Finally, the predicted sequence in the latent space can be decoded by the decoder module

| | $RMSE\ T$ (K) | $RMSE\ P$ (GPa) | $RMSE\ \bar{T}^{hs}$ (K) | $RMSE\ A^{hs}$ (µm²) | $RMSE\ \dot{\bar{T}}^{hs}$ (K/ns) | $RMSE\ \dot{A}^{hs}$ (µm²/ns) |
|---|---|---|---|---|---|---|
| PARCV1 | 157.18 | 1.342 | 378.86 | 0.0335 | 678.16 | 0.0524 |
| PARCV2 | 186.85 | 1.438 | 213.39 | 0.0324 | 527.62 | 0.0420 |
| LATENTPARC | 166.42 | 1.547 | 307.33 | 0.0234 | 688.05 | 0.0476 |

**Table 1: RMSE of sensitivity quantities of interest for PARCv1, PARCv2 and LatentPARC when compared to synthetic ground truth.**

simultaneously to make observations in the original state space.

### 3.4 Quantitative Metrics

To assess the predictive quality of LatentPARC when compared to PARCv1 and PARCv2 in a quantitative manner, we follow the methods detailed in [11]. We evaluate solution quality using the average hotspot temperature and hotspot area, consistent with standard practice in the energetic materials community [39]. These metrics capture both the intensity and spatial extent of hotspot growth, which are critical for assessing initiation behavior. We calculate the RMSE between model prediction and synthetic ground truth for six metrics: full temperature field, full pressure field, average temperature of the hotspot ($\bar{T}^{hs}$), hotspot area ($A^{hs}$), rate of change of hotspot temperature average ($\dot{\bar{T}}^{hs}$), and rate of change of hotspot area ($\dot{A}^{hs}$).

### 3.5 Data

The energetic material (EM) dataset was generated using the SCIMITAR3D code base [3], [4], [5], a numerical simulation method. The EM dataset consists of 134 instances of a single pore collapse. Each sample consists of 15 time-resolved snapshots of the temperature, pressure, and microstructure fields taken at uniform intervals of $\Delta t = 0.17\ ns$, capturing the pore collapse and hot spot formation of each sample. The spatial domain spans $1.5 \times 2.25\ \mu m$, with a planar shock of strength $P_s = 9.5\ GPa$ applied from left to right. The only varying initial condition across samples is the pore geometry, enabling focused study of morphology-driven dynamics under shock loading. Of the 134 total samples, 95 were used for training, 5 for validation, and 34 held out for testing.

# 4 Results

### 4.1 Predictive Quality of LatentPARC

To evaluate LatentPARC's predictive ability we compare our model to PARCv1 and PARCv2, two PARC models trained on the original data in the higher dimensional space. Each trained model is presented with the initial condition of a sample, unseen during training, which consists of the initial temperature, pressure, and microstructure fields. From this single snapshot, the resultant field evolution snapshots are predicted in an entirely auto regressive manner.

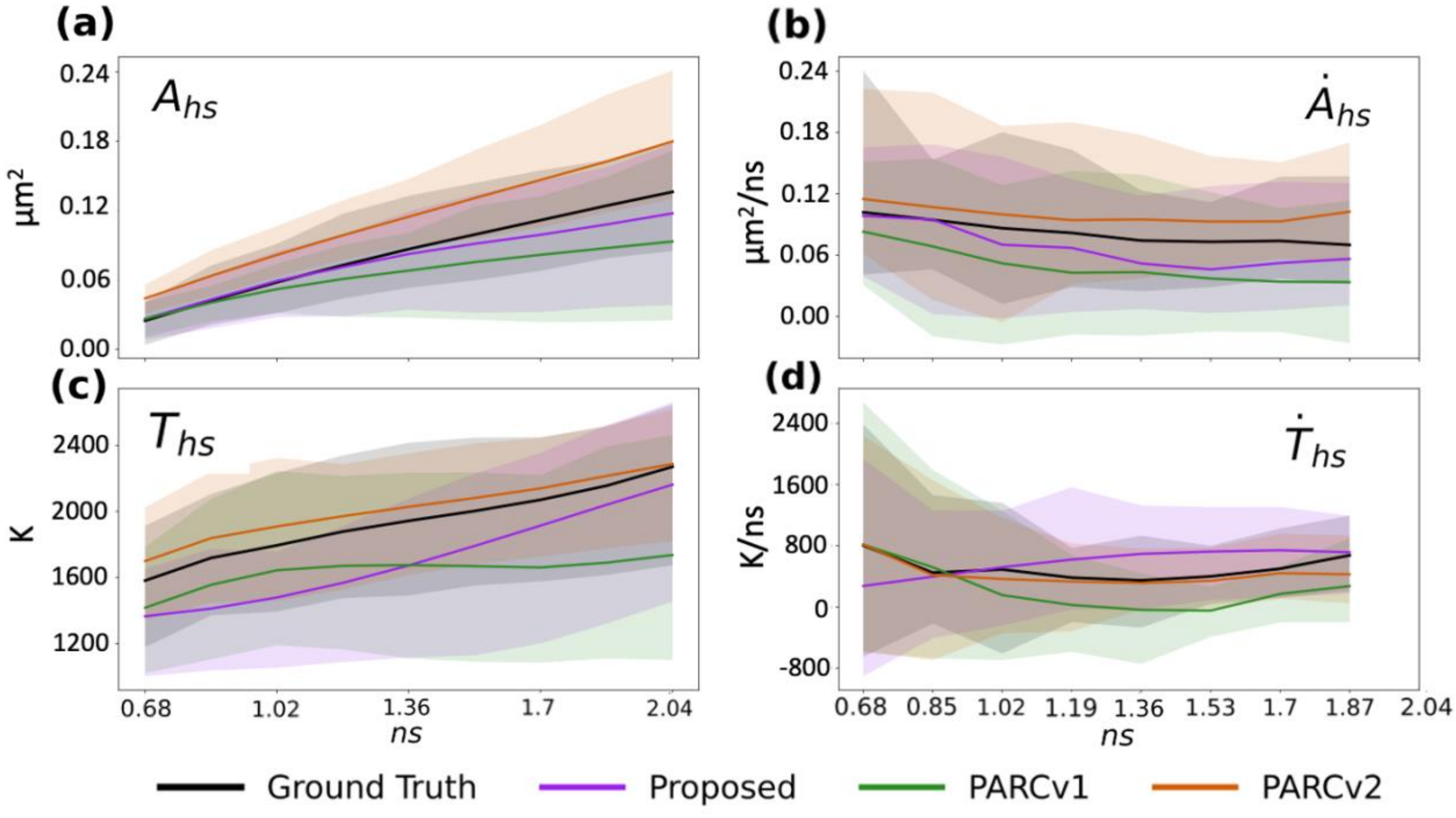

**Figure 5: Plots detailing the calculated quantity of interest values across time steps of the prediction.**

Table 1 summarizes the quantitative comparison between LatentPARC, PARCv1, and PARCv2 predictions. As shown, our model achieves an improvement upon PARCv1 in average hotspot temperature, hotspot area, and rate of change of hotspot area. In all other categories, LatentPARC performs similarly to PARCv1. However, the hotspot quantities of interest are of most importance here since they capture the accuracy of physical information that are essential to the usefulness of our model. PARCv2 remains state-of-the-art in all four hotspot metric categories.

We detail the trends of our four quantities of interest over the full time series of our model predictions in the plots in Figure 5. Here, we average the performance metrics across all 34 test samples and plot the mean and standard deviation. In these plots, the synthetic ground truth is plotted in black, and the closer the value of a model is to that line, the better the model does on that quantity of interest. LatentPARC performs the best of the three models on hotspot area and rate of change of hotspot area. LatentPARC performs better than PARCv1 on average hotspot temperature and about the same on average hotspot temperature rate of change.

**Figure 6: Comparison between temperature field predictions on a test sample.**

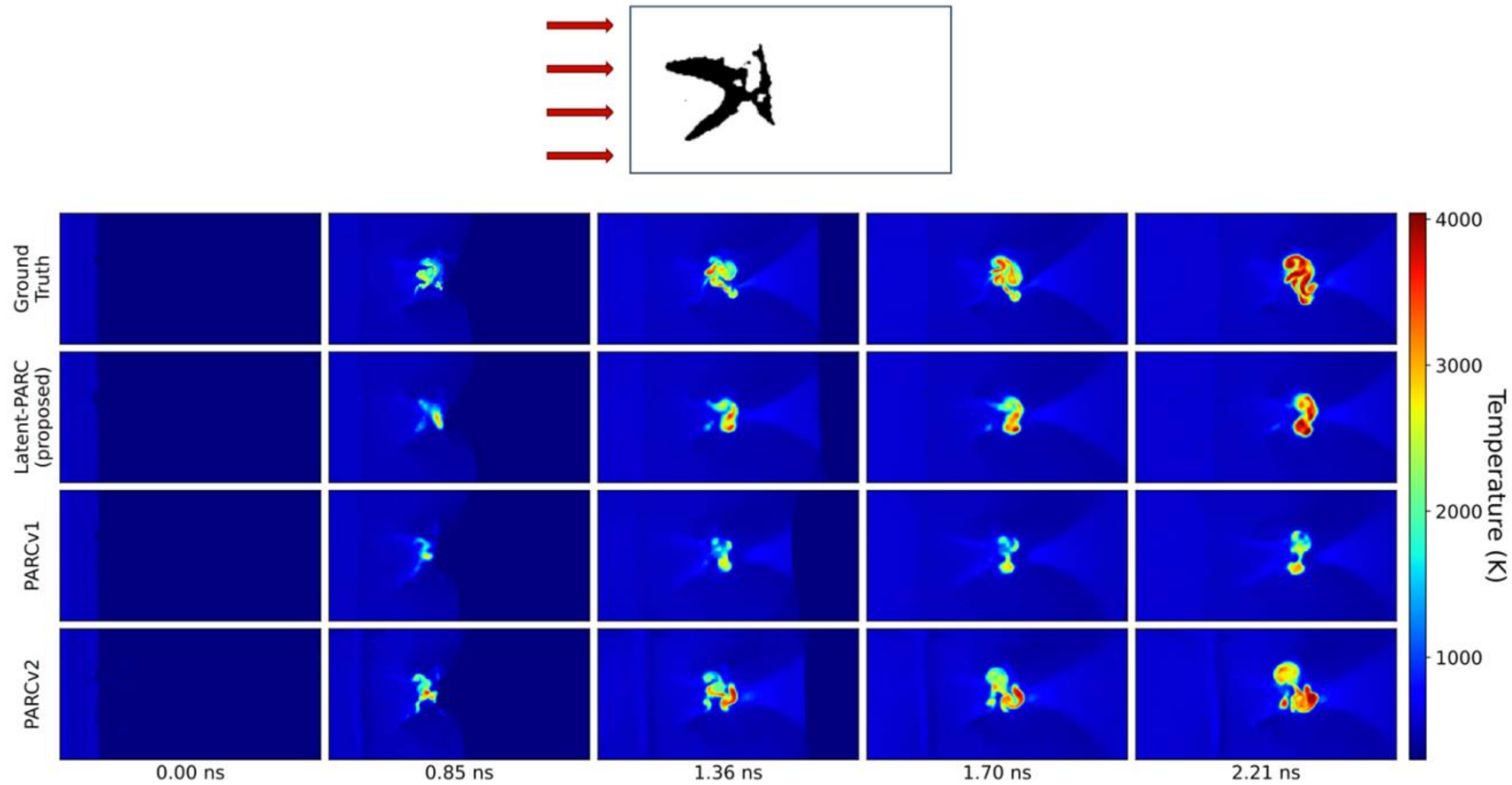

**Figure 7: Comparison between temperature field predictions on another test sample.**

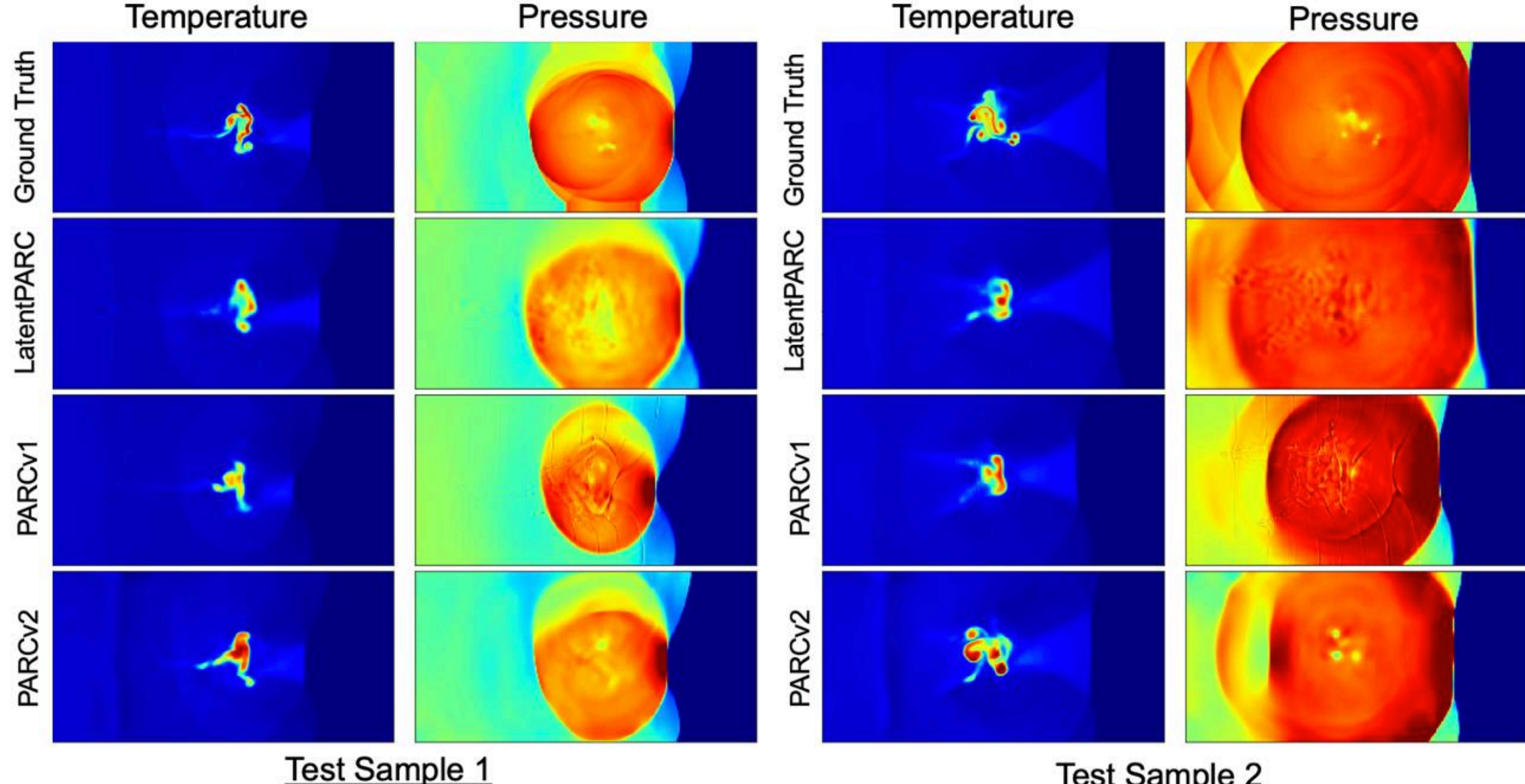

**Figure 8: Comparison between temperature and pressure field predictions on two test samples.**

Figures 6 and 7 demonstrate a qualitative comparison between LatentPARC, PARCv1, PARCv2, and numerical simulation data which we use as our synthetic ground truth. We observe, qualitatively, that LatentPARC captures key features of the reaction such as shock front, hotspot location, hotspot size, and general hotspot geometry with similar ability to PARCv1 and PARCv2. In some cases, LatentPARC captures details of the shock front better than both alternate models. We also observe that LatentPARC is more numerically stable than PARCv1 during later time steps. For example, PARCv1's prediction in the later time steps in Figure 6 show the beginning stages of error accumulation which result in a corrupted hotspot formation with exceedingly high temperatures. PARCv1 also shows a tendency to underpredict temperature magnitude in later timesteps as observed in Figure 7 and by the higher RMSE of average hotspot temperature, specified in Table 1.

However, LatentPARC fails to capture the fine-scale details of the hotspot geometry when compared to PARCv2 and tends to produce blurrier predictions than both PARCv1 and PARCv2. This is expected, both since LatentPARC's PARC module is more like PARCv1, which is also unable to capture these finer details, and due to the encoding process being slightly lossy. This discrepancy is acceptable for LatentPARC since it maintains low error in quantitative metrics shown in Table 1. However, implementing some version of PARCv2 in the latent space remains a future direction of interest. Since our focus in this study was on establishing the feasibility of a PARC model of the complex spatiotemporal dynamics of EM within a latent space, we feel this is suitable to leave for future work.

Figure 8 shows a snapshot of the predicted pressure fields in addition to temperature fields on two additional test samples. Here we observe similar trends in model performance on the temperature field predictions. We also, however, note some interesting details in the pressure predictions. We observe that the pressure field predictions of LatentPARC are blurrier than those of PARCv2 but also don't suffer from artifacts such as those observed in the PARCv1 predictions (vertical bands retained from shock fronts at previous time steps).

## 4.2 Computational Efficiency of LatentPARC

Once trained, LatentPARC has an inference speed which is roughly 10x and 30x faster than PARCv1 and PARCv2, respectively. In addition to the model speedup, LatentPARC greatly reduces the number of trainable parameters, and thus model size, when compared to previous full order PARC models. Details of these results are specified in Table 2.

Also of note, is a great reduction in the time it takes to train LatentPARC when compared to PARCv1 and PARCv2. We report the training time required for a single epoch of each model in Table 3 for rollout training where the window size is $2$ (snapshot prediction) and where the window size is $5$. Here, the same data, batch size of 1, hardware (A100 GPU) were used. Typically, these models train to around $3{,}000$ epochs. For the larger models, rollout training with a window of $5$ is very expensive, so it is used only for fine-tuning the model after it has converged on snapshot training. Since LatentPARC is much lighter and quicker to train, we can train the model entirely with a window of $5ts$.

| | *Prediction Speed (GPU)* | *Parameters (PARC Module)* | *Parameters (AE)* |
|---|---|---|---|
| PARCv1 | 0.50 s | 17,852,019 | X |
| PARCv2 | 1.6 s | 19,229,509 | X |
| Latent PARC | 0.045 s | 1,263,144 | 1,611 |

**Table 2: Comparison of computational efficiency between PARCv1, PARCv2, and LatentPARC.**

| | *2ts Window (minutes)* | *5ts Window (minutes)* | *Auto-encoder (minutes)* |
|---|---|---|---|
| PARCv1 | ~1.09 | ~1.95 | X |
| PARCv2 | ~6.41 | ~18.52 | X |
| LatentPARC | ~0.27 | ~0.62 | ~0.05 |

**Table 3: Comparison of training time between PARCv1, PARCv2, and LatentPARC.**

# 5 Discussion

In this work, we introduced LatentPARC, a reduced-order modeling framework for a PARC model that operates in a latent space and apply it to the problem of modeling EM thermomechanics. LatentPARC combines the high predictive fidelity of PARC models with greatly improved computational efficiency and achieves a reduced model size and accelerated training and inference speed relative to full-order PARC models. LatentPARC's reduced model size positions it as a model with greater flexibility and accessibility than PARCv1 and PARCv2, enabling deployment on resource-constrained hardware such as smaller GPUs without sacrificing the accuracy of PARC models. This opens a possibility to scale PARC to a larger scale simulation of EMs. The significant speed up in inference time when paired with this maintained accuracy makes LatentPARC an ideal tool for making an extensive exploration of the EM design space more feasible.

Given that LatentPARC has the simplest implementation of a PARC module within its latent space, making it most comparable to PARCv1, it is notable that our model demonstrates a marked improvement in predictive accuracy over PARCv1. This result supports our hypothesis that projecting our data onto a reduced-dimensional manifold yields dual benefits: (i) the latent representation provides a more compact and structured encoding of the training data, which simplifies the learning of dynamics and improves predictive fidelity; and (ii) the dimensionality reduction itself affords significant gains in computational efficiency and model flexibility. These findings suggest that learning dynamics in a reduced representation space can be advantageous for physics-informed machine learning, even in the absence of explicit constraints on the latent embedding. Importantly, the fact that performance improvements arise despite limited control over the learned latent representation highlights a promising future direction investigating how to learn a latent representation which is explicitly structured to be ideal for the learning of dynamics.

Beyond these immediate computational gains, the results of LatentPARC carry broader implications for the modeling of EMs and for ROM development more generally. The challenge of modeling EMs spans across multitudes of high-dimensional data at multiple scales with strong nonlinear behavior, making it a challenging problem for ROM. By demonstrating that PARC can be successfully projected into a latent space without significant loss in predictive fidelity, we establish that latent neural network–based ROMs are a viable pathway for handling EM thermo-mechanics. Because of this capability, LatentPARC opens the door to rapid simulation at larger scales—both in terms of grid resolution and number of time steps—and provides a foundation for extending PARC-style modeling to full 3D simulations. This advance not only accelerates computational exploration of EM phenomena but also suggests that similar latent-space methods may be extended to other domains with comparable complexity, such as turbulent flows, combustion, and multiphase materials systems.

While LatentPARC demonstrates strong results, several limitations should also be acknowledged. LatentPARC has not yet been tested on problems in other settings. To push LatentPARC further, we would like to test it on EM data with varying initial conditions as well as other physics problems. Applying LatentPARC to single pore EM first was done with the motivation of faster inference time being a priority for EM design exploration and because this is a difficult problem which serves as a convincing comparison to PARC models. Similarly, the current implementation can only explore a limited search space of single-pore geometries. Additionally, once LatentPARC is trained, it can only perform predictions within the initial conditions of its training data. While LatentPARC performs very well on unseen pore geometries, this constraint reduces the practicality of LatentPARC use across initial conditions and material formulations, although it is important to note that other PARC models and physics informed ML models face similar limitations. Extending LatentPARC to more diverse conditions would require retraining with more extensive data, which remains an area for future work.

# 6 Conclusion

In this work, we introduced LatentPARC, a reduced-order modeling framework that projects PARC into a latent space to achieve significant efficiency gains while preserving predictive fidelity. By reducing inference time and model size, LatentPARC enables broader and more accessible exploration of energetic material thermo-mechanics, lowering barriers to simulating resultant thermodynamics of complex microstructures thus accelerating EM design. While current limitations, such as fixed initial conditions, single-pore geometries, and an inability to take a PARCv2 approach in the latent space are present, our results demonstrate the feasibility of latent neural network–based ROMs for EM systems and highlight a stable training strategy for balancing reconstruction fidelity with accurate dynamics prediction. Beyond EMs, these findings suggest that latent ROMs offer a scalable pathway for modeling other highly nonlinear, multiscale systems, positioning LatentPARC as both a practical tool for materials scientists and a foundation for future advances in ML-driven reduced-order modeling.

## Acknowledgments

This work was performed under the auspices of the U.S. Department of Energy by the Lawrence Livermore National Laboratory under Contract DE-AC52-07NA27344 and was supported by the LLNL-LDRD Program under Project No. 24-SI-004. This work was also partially supported by the National Science Foundation under Grant No. DMREF-2203580. LLNL release number: LLNL-JRNL-2010773.